# Artificial ozone holes

S. N. Dolya

*Joint Institute for Nuclear Research, Joliot Curie str. 6, Dubna, Russia, 141980*

This article considers an opportunity of disinfecting a part of the Earth surface, occupying a large area of ~ $10^4$ km$^2$. Such a need may arise, for example, in the case of uncontrolled spreading out of different viruses or other infections. The method of disinfection being described in this article is more close to the nature than the radiation sterilization.

Decontamination of material and objects by means of this method will be carried out by using strong ultraviolet solar radiation through an artificial ozone hole in the Earth atmosphere. To realize this method, it is necessary to launch a balloon into the atmosphere, higher than the upper limit of the distribution of ozone into the atmosphere to the height of $H_1 = 30$ km. In the balloon gondola there is bromine in the solid state. The sunlight will cause dissociation of molecular bromine into atoms, each bromine atom kills $M_{Br + O3} = 3 * 10^4$ molecules of ozone. Each bromine plate has a mass of $4 * 10^{-2}$ grams and destroys ozone in the area of 10 x 10 meters. Thus, to form the ozone hole over the area of 10 thousand square kilometers, it is required to have the total mass of bromine equal to the following: $m_{Br\,1} = 4$ tons.

**Introduction**

The main source of ultraviolet radiation on the Earth is the Sun. The Sun radiation power is much stronger than all the existing light sources on the Earth. Although the power density of solar radiation in the hard ultraviolet is rather small [1], due to the fact that materials and objects can be irradiated on the area of thousands of square kilometers, the total power can be hundreds of Giga Watts, that is impossible to have with mercury lamps.

However, almost all hard ultraviolet radiation of the Sun is absorbed by the ozone layer of the Earth [2]. The natural ozone holes (thinning of the ozone layer), are dangerous because in the absence of ozone the hard ultraviolet radiation of the Sun penetrates through the Earth atmosphere and produces undesirable sterilization of materials and objects.

Ozone is effectively decomposed by bromine in the following chemical reactions:

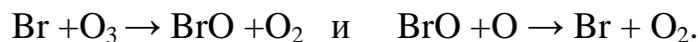

$$Br + O_3 \rightarrow BrO + O_2 \quad \text{и} \quad BrO + O \rightarrow Br + O_2.$$

In the result of these reactions, the molecules of ozone $O_3$ are converted into oxygen molecules $O_2$, but initial atoms of bromine remain in the free state and again participate in this process. Every atom of bromine destroys ozone molecules: according to [3], this number is equal to a million ($10^6$), the other data [4] give the number equal to $10^5$ ozone molecules. After that it escapes from the atmosphere in the result of other chemical reactions. Let us optimistically assume that the multiplicative parameter $M_{Br + O3}$, in this reaction is equal to: $M_{Br + O3} = 3 * 10^4$.

We remember that atomic bromine is obtained from dissociation of $Br_2$ under the influence of sunlight.

**1. The drift of bromine molecules in the air under the action of gravity**

We consider the process of bromine molecules coming down to the Earth under the action of gravity after their spraying out in the stratosphere above the region occupied by the ozone layer. Besides the gravity the molecules of bromine are influenced by the resistance related with their friction with the air.

We find the drift velocity from the following equation:

$$m_{Br} \, dV_z/dt = m_{Br} g - C_z S_{tr\,m} \rho V_z^2/2, \qquad (1)$$

where $m_{Br}$ is the mass of the molecule of bromine, $V_z$ - vertical velocity, $g = 10^3$ cm / s$^2$ is the acceleration of gravity, $C_z$ - drag coefficient, $S_{tr\,m}$ – transverse cross-section of the molecule of bromine.
The $\rho = \rho_0 * \exp[-z / H_0]$ – is a barometric formula, where
$\rho_0 = 1.3 * 10^{-3}$ g / cm$^3$ which is the air density under normal conditions,
$H_0 = 7$ km, it is the altitude at which the air density decreases by factor e.

Assuming the established velocity of bromine atoms to be equal to $dV_z / dt = 0$, we find from (1)

$$V_{z\,drift} = (2m_{Br}\, g/C_z S_{tr\,m} \rho)^{1/2}. \qquad (2)$$

After substituting the numerical values: $m_{Br} = 160 * 1.6 * 10^{-24}$ g, $C_z = 1$, $S_{tr} = 10^{-16}$ cm$^2$ that is the transverse cross-section of the molecule of bromine, $\rho = 2.6 * 10^{-5}$ g / cm$^3$, we find that the drift velocity, i.e. – the velocity at which the molecules of bromine are falling to the Earth from the height $H_1 = 30$ km, is: $V_{z\,drift} \approx 10$ cm / s.

**2. Diffusion of bromine molecules in the radial direction**

Consider the process of "spreading" the gaseous bromine cloud over the radius while it's moving slowly downwards under the influence of gravity. The cloud will increase its size due to diffusion.

The diffusion process will go on the radius according to Einstein's formula:

$$r^2 = D*t, \qquad (3)$$

where D is the diffusion coefficient for gases equal to: $D = (1/3) l_a * \bar{v}$, $l_a$ - the mean length of free passing, v is the average velocity of bromine molecules, r – the radial coordinate, t - time.

Define, first, the mean free passing by using the following formula:

$$l_a = 1/n\sigma, \qquad (4)$$

where n is the number of molecules of air in cubic centimeters at a given height, σ – their collision cross section.

According to the barometric formula, the number of molecules in a cubic centimeter at a height $H_1 = 30$ km, will be by 50 times less than the number Loschmidt, i.e., the number of molecules in a cubic centimeter under normal conditions. Taking into account the Loschmidt number equal to: $n_0 = 2.7 * 10^{19}$ molecules / $cm^3$, we find $n = 5.4 * 10^{17}$ molecules / $cm^3$. As the interaction cross section we can take the transverse cross-sectional of the molecule of bromine, which was taken to be equal to: $S_{tr\,m} = 10^{-16}$ $cm^2$. Then, from formula (4), $l_a = 2 * 10^{-2}$ cm.

Now we define the quantity $\bar{v}$ - the average velocity of bromine molecules. We find it from the following relation:

$$m\bar{v}^2/2 = (3/2)k*T, \qquad (5)$$

where $k = 1.38 * 10^{-16}$ egr / degree - Boltzmann constant, $T = 300\ ^0K$ - gas temperature. Then, according to the formula (5), the average velocity of the molecules of bromine is: $\bar{v} = 3 * 10^4$ cm / s.

You can now find $D = (1/3) l_a * \bar{v}$, - diffusion coefficient of bromine molecules, which in our case is equal to: $D = 2 * 10^2 \text{ cm}^2 / \text{s}$.

The average radius of the spreading out the bromine molecules in space will grow with time as:

$$r = (D*t)^{1/2}. \qquad (6)$$

It can be seen that the characteristic time: $t_{10} = 10^5$ s, during which the bromine molecules will drift down from the height of $H_1 = 30$ km at a distance $\Delta H_1 = 10$ km, along the radius the bromine cloud will spread out at a distance of: $r_{10} = (D * t)^{1/2} = 40$ meters.

This means that if we want to uniformly fill a large area (100*100 km), then the initial bromine clouds must be placed from each other at a distance of not more than $\Delta r_{in} < 10$ meters.

## 3. Uniform distribution of crystallites of bromine over a large area

We consider how to uniformly fill a large area with gaseous bromine. We take the case when the bromine presented in the form of small solid crystals is thrown out by means of a centrifuge, similar to the way as the road machines throw sand onto the roadway.

The bromine crystals should fly quite a long distance in the radial direction - about 50 kilometers before reaching the desired point. Having reached it they have to melt, warm up till the evaporation temperature and evaporate. The distance between the neighboring crystals should be not more than 10 meters for after their evaporation all the required area would be uniformly filled with gaseous bromine.

### *3.1. Quantitative relationships*

In the column with an area of 1 cm$^2$ at normal conditions there are $10^{19}$ ozone molecules, in the column with an area of 1 m$^2$ their number is $10^{23}$. We consider the processes in the standard "box" with dimensions of 10*10 m, in the column with this window there are $10^{25}$ molecules of ozone.

We assume that one bromine atom decomposes $3 * 10^4$ molecules of ozone. It means that in order to decompose all the molecules in this window it will be required to have $3 * 10^{20}$ atoms of bromine.

At 100% dissociation in the column there must be $1.5 * 10^{20}$ molecules of bromine, taking into account that the bromine molecule consists of two atoms Br.

Let us find the mass of molecular bromine from the relation:

$$6*10^{23} \text{ ------------ } 160 \text{ g}$$
$$1.5*10^{20} \text{ ------------ } x \text{ g},$$

from where $x = 4 * 10^{-2}$ g, i.e. the mass of a crystal of bromine is equal to $m_{Br} = 4 * 10^{-2}$ g. Densities of solid and liquid bromine differ by less than 1%, [6], p. 188. The density of liquid bromine is: $\rho_{Br} = 3.12$ g / cm$^3$, then the volume occupied by a drop of bromine is equal to: $V_{Br} = 10^{-2}$ cm$^3$.

It is possible to find the ball diameter, assuming that the volume of the ball: $V_{Br} = (4/3) \pi r_{Br}^3 = (\pi / 6) d_{Br}^3 \approx d_b^3 / 2$. From this we find: $d_{Br} = 0.27$ cm. The ball of this diameter contains $1.5 * 10^{20}$ molecules of bromine, which, after dissociation will give $3 * 10^{20}$ bromine atoms and these atoms will be able to kill $10^{25}$ ozone molecules in a column with a base of 10*10 m.

*3.2. Aerodynamics of bromine crystals having a spherical shape*

We find the Reynolds number for a ball of solid bromine (the melting temperature of bromine is $T_m = -7.3$ $^0$C) according to the following formula:

$$Re = \rho V_0 l / \eta, \qquad (7)$$

where $\rho$ is the density of air at a given altitude, 0 - the velocity of the ball in the air, l - the characteristic length, in our case: $l = d_{Br} = 0.27$ cm,
$\eta = 1.8 * 10^{-4}$ Poise - the viscosity of air, [5], p. 273. Substituting numbers for the height $H_1 = 30$ km, and for the initial velocity of the ball $V_0 = 200$ m / s, we find that the Reynolds number: $Re = 7 * 10^2$. This means that the dependence of the resistance on the velocity is linear and the velocity in dependence on the distance will decrease exponentially:

$$V = V_0 \exp[-(\rho C_x S_{tr}/m_{Br})x], \qquad (8)$$

where $C_x$ – the drag coefficient of the ball, $C_x = 0.6$, $S_{tr}$ – the transverse cross-section of the ball $S_{tr} = \pi d_{Br}^2 / 4 = 5.7 * 10^{-2}$ cm$^2$.

The value shown in the parentheses of the exponent, the so-called "ballistic coefficient" describes the "quality" of the object. The value inverse to the ballistic coefficient has the dimension of length and can be considered as the characteristic length on which the body stops since its velocity decreases by factor e. In our case, the length is equal to:
$x_0 = m_{Br} / \rho C_x S_{tr} = 4 * 10^{-2} / (2.6 * 10^{-5} * 0.6 * 5.7 * 10^{-2}) = 4.6 * 10^4$ cm, or $x_0 \approx 500$ m.

### *3.3. Possible sizes of the ozone hole and the required mass of bromine for its forming*

Suppose you want to form the ozone hole with the dimensions of 100*100 km. This area contains $10^8$ "windows" with the size of 10 * 10 m, in each of them it is necessary to place a crystal of bromine with a mass of $m_{Br} = 4 * 10^{-2}$ g.

Thus, to form the ozone hole with dimensions of 100*100 km, it is necessary to uniformly distribute $m_{Br\ 1} = 4$ tons bromine over this area.

### *3.4. Opportunities of the balloon*

We consider what load capacity to a height of $H_1 = 30$ km has a balloon. Let the diameter of the ball be $d_b = 100$ m. Then, the total volume of such a balloon will be as follows: $V_b = (4/3) \pi r_b^3 = (\pi / 6) d_b^3 \approx d_b^3 / 2 = 5 * 10^5$ m$^3$.

The air density at the height of $H_1 = 30$ km, is $\rho = 2.6 * 10^{-5}$ t/ m3, so that the load capacity of the empty balloon is equal to13 tons. If the balloon is filled with hydrogen gas, whose density is by14 times less than the density of the air, the balloon load capacity will be by 7% lower than the empty balloon and equal to 12 tons.

Suppose that the balloon shell is made of a polymer film having the density of $\rho_b = 1.2$ g / cm$^3$, and thickness of $\delta_b = 25$ μ. Then 1 m$^2$ of such a film has a mass: $m_b = 30$ g. The area of the shell of the balloon is equal to:
$S_b = \pi d_b^2 = 3.14 * 10^4$ m$^2$. The mass of the shell is equal to $m_b * S_b = 1$ ton.

Thus, the load capacity of such a balloon is equal to 11 tons; 4 tons of them are the bromine mass. The remaining mass may consist of auxiliary equipment: a cooling chamber, the centrifuge needed for spreading out the crystals of bromine, energy supply, navigation, control, and so on.

*3.5. Aerodynamics of the bromine crystals having the form of thin plates*

It follows from (8), that the range of the flight of the bromine crystals in the Earth atmosphere at a height of $H_1 = 30$ km, is not great and is about 500 m. For uniform distribution of the bromine crystals on the surface with dimensions of 100 * 100 km, the flight distance must be greater than 50 km.

This distance of throwing out the bromine crystals can be obtained if they are formed as cylinders having a sharp cone at the head end.

Indeed, the velocity of bromine crystals chosen by us: $V_0 = 200$ m / s, is $M = 0.6$ Mach units, where $M = 1$ corresponds to the velocity equal to the velocity of the sound in air under normal conditions: $V_s = 330$ m / s. For a ball, for this velocity, the drag coefficient is $C_x = 0.6$, [6].

For a cylinder with a sharp cone in the head, according to the empirical formula [7], the drag coefficient is equal to:

$$C_x = (1.56 + 1.95/M^2)\, \Theta_{1/2}^{1.7}, \qquad (9)$$

where $M = 0.6$ - Mach number, $\Theta$ - the half-angle of the cone at the vertex. As the full angle at the vertex of the cone, it is possible to obtain the relation: $C_x = \Theta^2$, Newton's formula. It is seen that at a small angle of the cone, the drag coefficient may be very small.

To have the distance of throwing the crystals of bromine to be about 50 kilometers, according to the formula (8) the drag coefficient should be equal to: $C_x = 5 * 10^{-3}$, or the angle at the vertex of the cone $\Theta_s = 7 * 10^{-2}$.

If we imagine a bromine crystal as a cylinder with a mass $m_{Br} = 4 * 10^{-2}$ g, having a volume: $V_{Br} = 10^{-2}$ cm$^3$, its total length is equal to: $l_t = 3$ cm, diameter $d_{Br\,1} = 0.7$ mm and length of the cone part $l_c = 1$ cm. The angle at the vertex of the cone $\Theta_s = d_{Br\,1} / l_c$ if to substitute the numbers it will be equal to what we expected: $\Theta_s = 7 * 10^{-2}$.

*3.6. Phase transitions: solid body-liquid and liquid - vapor*

The bromine crystal must not only reach the given point, but it must reach it in a gaseous state. For bromine the phase transition energy from solid to liquid is $\Delta Q_s = 10$ kJ / mole, [8], 289.

Assuming 1 mole to be equal to 160 g, we find that the energy required to melt the crystal of mass $m_{Br} = 4 * 10^{-2}$ g, is: $\Delta Q_{m\,c} = 2.4$ J.

We find the value of energy transferred from the Sun to the crystals of bromine, assuming that the density of the solar energy is $P = 1.4$ kW / m$^2$. The area of the longitudinal cross section of the cylinder with a conical head is $S_{long} = d_{Br\,1} * l_t = 0.07 * 3 \approx 0.02$ cm$^2$. The flux density of the solar energy incident on a crystal of bromine is then equal to: $p_{Br} = 0.03$ W. It can be seen that during the flight of the crystallite $\tau_{Br} = 100$ seconds the energy required for melting the bromine crystal can be obtained from the Sun.

Besides the fact that it is necessary to melt the crystal of bromine, it has to be heated to the temperature of evaporation: $T_w = 59.2$ $^0$C and transferred to the vapor state. The energy required for the phase transition of bromine from liquid to vapor is $\Delta Q_{w\,c} = 29.5$ kJ / mole, i. e., it is about 3 times greater than the energy required for the phase transition of the solid body into the liquid state. This means that it is necessary to have the bromine crystals in the form of plates, i.e.: it is needed to develop the surface, through which the bromine crystals obtain the solar heat.

Thus, the bromine plate with a thickness of $\delta_{Br} = 7\mu$, the length of the conical part $l_{c\,p} = 0.1$ mm, and with the sizes $l_{Br} = 3.7$ x $3.7$ cm, has the appropriate drag coefficient, and the plate area is 700 times greater than the area of the longitudinal cross section of the cylinder.

*3.7. Slow down of bromine solid plates related with air viscosity*

Since the required form of the bromine crystals is the form of thin plates with a big lateral surface, then in the subsonic velocities the bromine crystals slowing down may be very strong due to the viscosity of the air.

The equation of motion of a solid bromine plate in the air can be written as follows:

$$m_{Br}\, dV_x/dt = -F_{x2}. \tag{10}$$

the force of the resistance related with viscosity:

$$F_{x2} = \sigma S = 2\eta(dV/dz)*l_{Br}^2 = 2\eta(V/\delta)*l_{Br}^2, \tag{11}$$

where $\eta = 2 * 10^{-4}$ Poise is the viscosity of air, $l_{Br} = 3.7$ cm - side of the plate square, $dV/dz$ - radial gradient of the longitudinal air velocity near the plate, $\delta$ – the characteristic length of velocity changing along the radius.

Equation (10) having the viscous force of slowing down of the bromine plate in the form of (11) can be transformed into:

$$dV/V = - (2\eta/m_{Br} \delta)*l_{Br}^2*dt, \qquad (12)$$

which has the following solution:

$$V = V_0 \exp[-(2\eta/m_{Br} \delta)*l_{Br}^2*t]. \qquad (13)$$

Now we can find the length being passed by the plate before its stopping in the air:

$$l_{air} = \int_0^\infty V dt = V_0(2m_{Br} \delta/\eta * l_{Br}^2). \qquad (14)$$

To find the length of the above length, it is necessary to find $\delta$ which is a characteristic length of the air velocity changing near the plate.

Find it from Navies equation:

$$\partial V_x/\partial t + V_x \partial V_x/\partial x = (\eta/\rho) \partial^2 V_x/\partial z^2 + (\eta/\rho)\partial^2 V_x/\partial x^2. \qquad (15)$$

Let us see in which case the term $V_x \partial V_x / \partial x$ will be much bigger than $(\eta / \rho) \partial^2 V_x /\partial x^2$. Let $\partial x = l_{ch}$ –that is a characteristic length at which the velocity changes. Then $V_x \partial V_x / \partial x = V_x^2 / l_{ch}$, $(\eta / \rho) \partial^2 V_x / \partial x^2 = (\eta / \rho) V_x / l_{ch}^2$. To satisfy the condition: $V_x \partial V_x / \partial x >> (\eta / \rho) \partial^2 V_x / \partial x^2$, it is necessary to fulfill $\rho V_x l_{ch} / \eta = Re >> 1$, where Re is the Reynolds number, and the ratio of Re $>>1$ in our case is certainly satisfied.

We now transform equation (15) to the following form:

$$\partial V_x/\partial t + V_x \partial V_x/\partial x = dV_x/dt = (\eta/\rho) d^2V_x/dz^2, \qquad (16)$$

where we have replaced $\partial / \partial t + V_x \partial / \partial x$ – a partial derivative with respect to time, with coordinate $d / dt$ (the total time derivative).

Substituting the above instead of the expression for the Navies $dV_x/dt$, determined from the equation of motion (12),
$dV_x/dt = -(2\eta/m_{Br}) * l_{Br}^2 * (dV_x/dz)$, we obtain the following equation:

$$-(2\eta/m_{Br})*l_{Br}^2*(dV_x/dz) = (\eta/\rho)\, d^2V_x/dz^2 , \qquad (17)$$

which, after reduction by $\eta$ - the viscosity of air and the above transfers, this equation changes to the following:

$$d^2V_x/dz^2 = -(2\rho/m_{Br})*l_{Br}^2*(dV_x/dz). \qquad (18)$$

Replacing $dV_x/dz$ for y we obtain:

$$d(\ln y) = -(2l_{Br}^2 \rho/m_{Br})dz, \qquad (19)$$

or

$$dV_x/dz = C_{1v}\exp(-2\rho l_{Br}^2 z/m_{Br}). \qquad (20)$$

From formula (20) it is seen that the characteristic length $\delta$, that is the change of longitudinal velocity $V_x$ in the transverse direction z, is equal to:

$$\delta = m_{Br}/2\rho l_{Br}^2. \qquad (21)$$

According to the formula (14) the slow down length of the plate in the air is equal to:

$$l_{air} = \int_0^\infty V dt = V_0(m_{Br}\delta/\eta*l_{Br}^2) = V_0\,[m_{Br}^2/\rho\eta l_{Br}^4]. \qquad (22)$$

Substituting numbers into (22): $V_0 = 2*10^4$ cm/s, $m_{Br} = 4*10^{-2}$ g, $\eta = 2*10^{-4}$ Poise, $l_{Br} = 3.7$ cm, we find the following:

$$l_{air} = 3.2*10^7 \text{ cm} = 320 \text{ km}. \qquad (23)$$

It can be seen that the plate slow down caused by the viscosity of air is not important.

The estimation of the vertical velocity of the plate displacement, obtained by formula (2), has shown that this velocity is small. During the flight time of the plate over the radius of 50 kilometers, it will slightly shift in the vertical

## 4. Operation of the equipment complex

The complex of equipment is shown in Fig. The balloon (1) raises a solid bromine gondola (2) to the required height. In the gondola there is also the appropriate cooling equipment, a centrifuge, navigation elements and power supply. The bromine plates (3) are thrown out from the gondola by means of the centrifuge over a large area. The balloon is raised to the height greater than the upper limit of the distribution of ozone in the atmosphere, shown by the curve (4). When in the atmosphere an artificial ozone hole is formed, the hard solar ultraviolet radiation reaches the Earth surface (5) and performs the required disinfection.

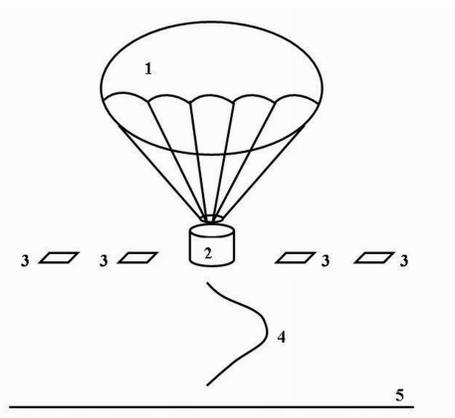

## Conclusion

The results of this work have shown that it is possible to make the crystals reach the given point in the gaseous state by using the proper choice of the form and sizes of the bromine crystals.

The solar radiation power in the ultraviolet wavelength range can be estimated as 50 W / $m^2$. Thus, that the total power per area of 100 * 100 kilometers, is equal to 500 Giga Watts that is absolutely not feasible for the devices located on the surface of the Earth.